\documentclass[useAMS]{mnras}
\usepackage[usenames,dvips]{color}
\usepackage{graphicx, color}
\usepackage{amssymb}
\usepackage{epsfig}

\newcommand{\chandra}{{\it Chandra}}

\newcommand{\swift}{{\it Swift}}
\newcommand{\rxte}{{\it RXTE}}

\newcommand{\ulx}{M82 X--1}

\newcommand{\flux}{\thinspace\hbox{$\hbox{ergs}\thinspace\hbox{cm}^{-2}\thinspace\hbox{s}^{-1}$}}

\def\spose#1{\hbox to 0pt{#1\hss}}
\def\laeq{\mathrel{\spose{\lower 3pt\hbox{$\mathchar"218$}} \raise 2.0pt\hbox{$\mathchar"13C$}}}
\def\gaeq{\mathrel{\spose{\lower 3pt\hbox{$\mathchar"218$}} \raise 2.0pt\hbox{$\mathchar"13E$}}}

\title[A Possible 55-day X-ray Period of M82 X--2]{A Possible 55-day X-ray Period of the Ultraluminous Accreting Pulsar M82 X--2}
\author[A.~K.~H.~Kong et al.]{Albert~K.~H.~Kong$^1$\thanks{E-mail: akong@phys.nthu.edu.tw}, Chin-Ping Hu$^2$, Lupin Chun-Che Lin$^3$, K.~L. Li$^4$, Ruolan Jin$^1$, 
\newauthor C.~Y.~Liu$^1$ and David Chien-Chang Yen$^5$\\
$^1$Institute of Astronomy and Department of Physics, National Tsing Hua University, Hsinchu 30013, Taiwan\\
$^2$Department of Physics, University of Hong Kong, Pokfulam Road, Hong Kong\\
$^3$Institute of Astronomy and Astrophysics, Academia Sinica, Taipei 10617, Taiwan\\
$^4$Department of Physics and Astronomy, Michigan State University, East Lansing, MI 48824-2320, USA\\
$^5$Department of Mathematics, Fu Jen Catholic University, New Taipei City 24205, Taiwan
}

\begin{document}

\pagerange{\pageref{firstpage}--\pageref{lastpage}}
\maketitle

\label{firstpage}

\begin{abstract} 
We report a possible detection of a 55-day X-ray modulation for the ultraluminous accreting pulsar M82 X--2 from archival \chandra\ observations. Because M82 X--2 is known to have a 2.5-day orbital period, if the 55-day period is real, it will be the superorbital period of the system. We also investigated variabilities of other three nearby ultraluminous X-ray sources in the central region of M82 with the \chandra\ data and did not find any evidence of periodicities. Furthermore, we re-examined the previously reported 62-day periodicity near the central region of M82 by performing a systematic timing study with all the archival {\it Rossi X-Ray Timing Explorer} and \swift\ data. Using various dynamic timing analysis methods, we confirmed that the 62-day period is not stable, suggesting that it is not the orbital period of M82 X--1 in agreement with previous work.
\end{abstract}


\section{Introduction} 
At the centre of the starburst galaxy M82, there are four interesting ultraluminous X-ray sources (ULXs; $L_X > 10^{39}$ erg s$^{-1}$).  The physical nature of ULXs is still in debate but it is now believed that ULXs have different types of populations. In the extreme end, some ULXs are very likely long-sought intermediate-mass black holes (e.g., Farrell et al. 2009; Pasham et al. 2014; Mezcua et al. 2015).  For instance, \ulx\ is one of the most promising sources hosting an intermediate-mass black hole with a mass of about 400 $M_\odot$ (Pasham et al. 2014).
The bulk of ULXs, however, can be explained by using stellar-mass black holes accreting at or above the Eddington limit (e.g., Gladstone et al. 2009; Motch et al. 2014). Furthermore, it has been proposed that the mass of the black holes in ULXs may be in the range of 20--30 $M_\odot$ (e.g., Liu et al. 2013), leading to a possible connection to the recent gravitational wave event detected by LIGO (Abbott et al. 2016). In addition to the population of black holes, the ULX M82 X--2 is recently confirmed as a neutron star system (Bachetti et al. 2014). Furthermore, some young X-ray supernova remnants can also be ultraluminous and M82 X--4 is one of the examples (Kong et al. 2007).

Motivated by the X-ray variability of \ulx, the centre of M82 has been monitored by several X-ray missions even though all the ULXs in the region are not well resolved with most of the instruments. One remarkable discovery is the 62-day X-ray periodicity by using \rxte\ (Kaaret et al. 2006; Kaaret \& Feng 2007) and it is suggested as the orbital modulation of \ulx. Subsequent analysis of more \rxte\ data reveals a phase shift in two different parts of the light curve (Pasham \& Strohmayer 2013) and a precessing accretion disc scenario is more likely. More recently, Qiu et al. (2015) show that by using \swift/X-ray Telescope (XRT) data, the 62-day periodicity is likely from a collection of periods of several luminous X-ray sources next to \ulx, which is not resolved by \rxte.

In this paper, we investigate the nature of the long-term X-ray modulations at the centre of M82. We obtained data from 25 \chandra\ observations distributed among 16 years to investigate the four brightest ULXs near the central region of M82. Comparing with the \chandra\ observations, we have more samples collected from \rxte\ and \swift\ observations to apply various dynamic timing analysis techniques for a detailed study in an expense of the spatial resolution to resolve our targets.

\section{Observations and Data Reduction}

\subsection{\chandra}
\chandra\ has observed M82 25 times from 1999 to 2015. Four observations were done with the High Resolution Camera (HRC-I or HRC-S) and others were taken with the Advanced CCD Imaging Spectrometer array (ACIS-I or ACIS-S). All the available data sets were reprocessed by using \texttt{CIAO} (version 4.7) and \texttt{CALDB} (version 4.6.8). We extracted the four targets X--1, X--2, X--3 and X--4 with elliptical source regions and nearby source free regions as backgrounds. Because the pointing of each observation is different, in some cases, the observations were highly off-axis so that some sources (usually X--2, X--3, and X--4) were not well resolved. We discarded them to minimise contamination. In this analysis, we have 29, 12, 21, and 23 data points for X--1, X--2, X--3, and X--4, respectively. Note that X--2 is a transient (Kong et al. 2007) and therefore it has fewer data points. Furthermore, we used slightly different source extraction regions in each observation to avoid contaminations by nearby sources. In general, we used different semi-major and semi-minor axes for different observations, and the range is 1.2\arcsec--3.3\arcsec, 0.5\arcsec--1.4\arcsec, 0.7\arcsec--2.3\arcsec, and 0.6\arcsec--1.5\arcsec\ for X--1, X--2, X--3, and X--4, respectively. Since we need to construct the long-term light curves by using all the ACIS and HRC data, we converted the instrumental count rate into flux. By using the \texttt{srcflux} tool in CIAO and assuming an absorbed power-law model with a photon index of 1.7 and N$_{H}$ = 3$\times$ 10$^{22}$ cm$^{-2}$ (adopted from Chiang $\&$ Kong, 2011), we generated the absorbed fluxes in 0.3--10.0 keV of all the ACIS data. For the HRC observations, we converted the count rate of each source into fluxes with PIMMS by applying the same spectral parameters used in ACIS data sets. 

\subsection{\rxte}
We used all the \rxte\ Proportional Counter Array (PCA) data of M82 taken between 1997 and 2009 in this study. Because of the poor spatial resolution of \rxte/PCA, the X-ray emission of M82 as seen by PCA is dominated by \ulx\ with contribution from nearby ULXs (see e.g., Kong et al. 2007).
We extracted the pipeline produced Standard-2 background-subtracted 2--9 keV light curves. There are altogether 857 valid light curves and we computed the average count rate for each observation to obtain the long-term light curve.

\subsection{\swift}
We used all the 180 \swift/XRT observations in photon counting mode taken from MJD 56022 (2012 April 05) to MJD 57053 (2015 January 31) throughout the analysis. As a huge X-ray brightening (i.e., 4 times larger than the usual) has been detected from M82 (and probably from M82 X--1) since 2015 January (i.e., the last data of this study), we skipped these observations to avoid contaminations from the unknown X-ray activity. Unlike Qiu et al. (2015) that excluded 70 observations in 2014 to avoid a possible contamination from the Type Ia supernova SN 2014J, our sample includes all these data based on the fact that no X-ray signal is detected with a 47~ks deep \chandra\ observation, with which the $3\sigma$ upper limit is $2.6\times10^{-15}$\flux\ (0.3--10~keV; Margutti et al. 2014). All the data were reprocessed by \texttt{xrtpipeline} of \texttt{HEAsoft} version 6.17, with updated \texttt{CALDB} files. Light curves with an energy range from 0.3 to 10~keV were extracted by \texttt{xrtgrblc} with 18\arcsec\ or 4\arcsec\ radius circular regions centred at M82 X--1 (with 4\arcsec\ and 18\arcsec), X--2 (with 4\arcsec), X--3 (with 4\arcsec), and X--4 (with 4\arcsec). All the source positions are according to Chiang \& Kong (2011). A 47\arcsec\ source-free background region was chosen close to the ULXs enough to estimate the background counts well, while also far enough to avoid the X-ray diffuse X-rays of M82. Note that all the extraction sizes adopted are the same as the ones described in Qiu et al. (2015) for a fair comparison.

\section{Data Analysis and Results}
\begin{figure}
\epsfig{file=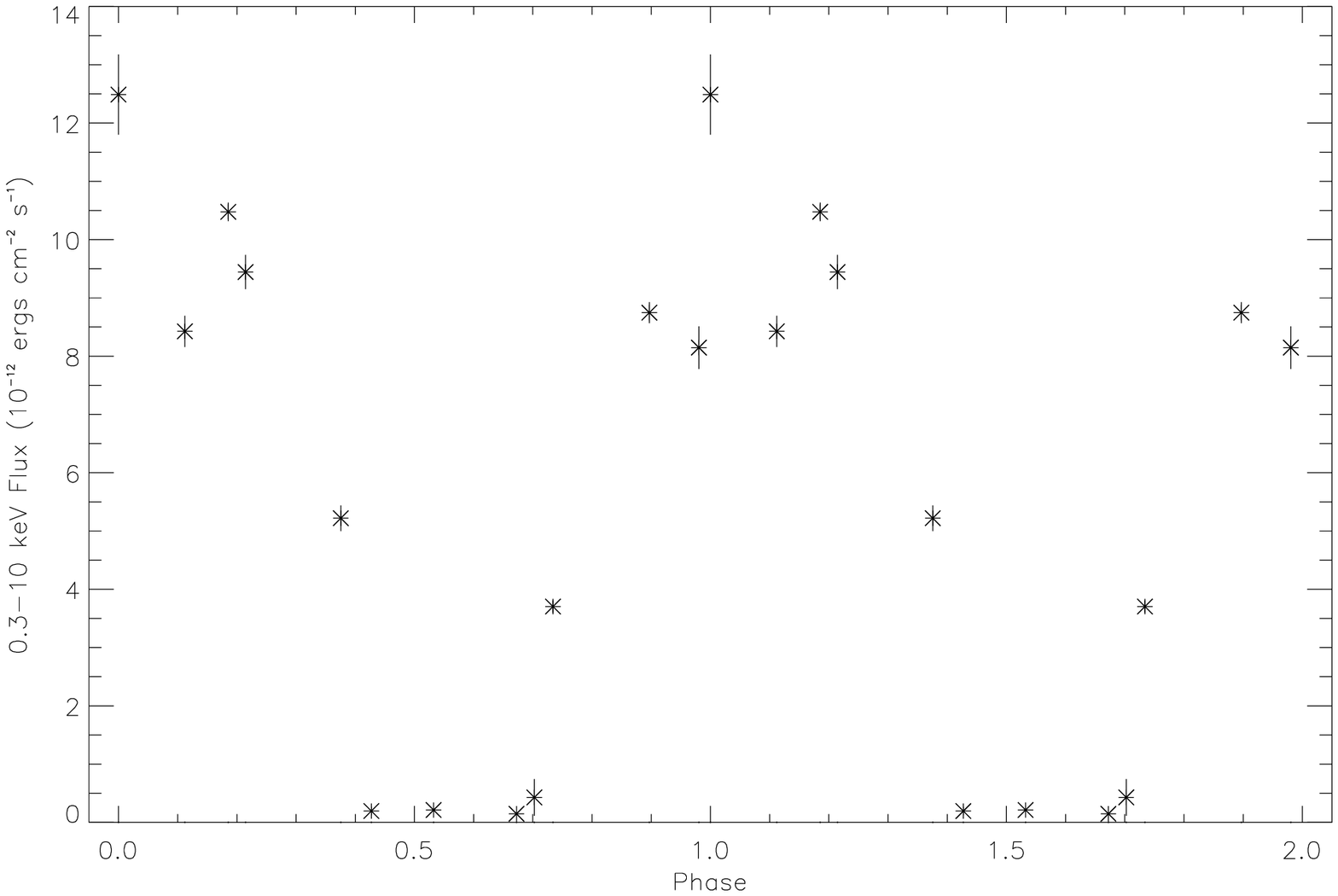,width=3.3in}
\epsfig{file=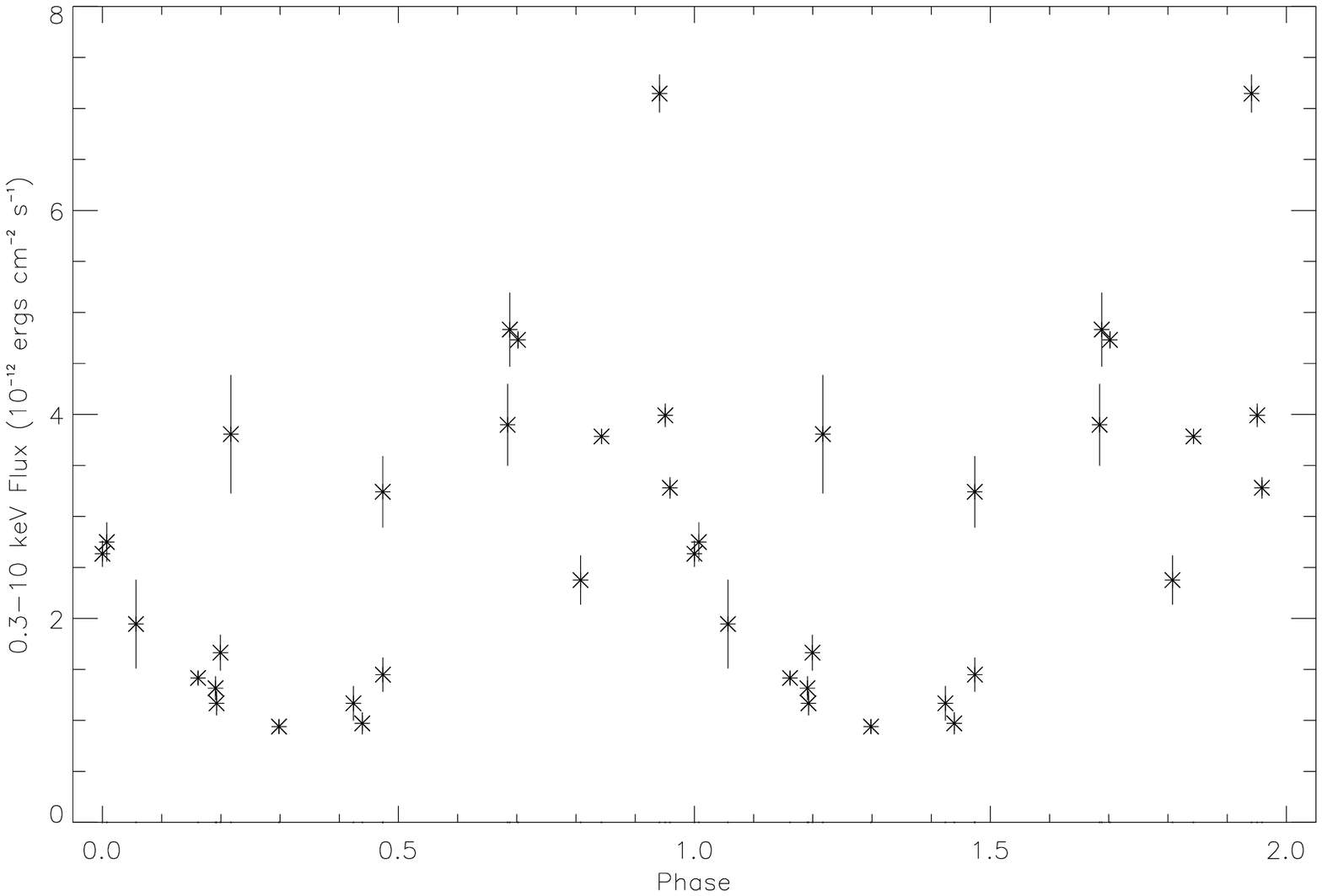,width=3.3in}
\caption{Folded \chandra\ light curves for M82 X--2 (upper panel) and X--3 (lower panel) with a period of 55.5 and 55.1 days, respectively. \label{fold_55d}}
\end{figure}

In Qiu et al. (2015), they suggested that the 62-day X-ray period seen in \rxte\ and \swift\ is likely to be a combination of different signals from several luminous X-ray sources in the region. Since the spatial resolution of \rxte\ and \swift\ is not sufficient to resolve all the sources, we here considered \chandra\ (both ACIS and HRC detectors) observations which can provide much better spatial resolution to investigate any quasi-periodic signal emerging from M82 X--1 to X--4.
Owing to the sparse data points over the last 16 years, we folded the data for all ULXs of M82 to examine whether there exists any periodicities (in the range of $60.5-62.5$ days, $54.5-56.5$ days, and $46.5-48.5$ days) consistent with the signals detected by \swift\ data (see the 4th paragraph of this section for details), and found that M82 X--2 and X--3 demonstrate a possible detection at $\sim 55$ days.
The corresponding folded light curves are shown in Figure \ref{fold_55d}.  
By performing $10^7$ Monte-Carlo simulations, the false alarm probability to get a better $\chi^2$ via a sinusoidal fitting is $\sim 0.00114$ and $\sim 0.078$ for M82 X--2 and X--3, respectively. Hence, we can reject M82 X--3 statistically for the 55-day period.

While the number of \chandra\ observations is limited, we would also like to investigate carefully if previous \rxte\ and \swift\ data may provide some hints on all the suggested signals.
We first produced the Lomb-Scargle periodogram (LSP; Lomb 1976; Scargle 1982) for \rxte\ and \swift\ data. For \rxte\ data, we obtained a significant signal at about 62 days ($\sim 0.01613$ 1/day) as in Kaaret \& Feng (2007). We also divided the dataset into two segments and found that the period changes slightly as indicated in Pasham \& Strohmayer (2013). To study this time-dependent behaviour, we examined the dynamic power spectrum (DPS, Clarkson et al. 2003), the weighted wavelet $z$-transform (WWZ, Foster 1996), and the Hilbert-Huang transform (HHT, Huang et al. 1998). The technical details of all these methods on the study of long-term X-ray variability are discussed in Lin et al.~(2015) and Hu et al.~(2014).  Only those data points after $\sim$MJD 53,800 were included in the time-frequency analysis because we would like to avoid any artificial signals originated from the large data gaps.

\begin{figure}
\epsfig{file=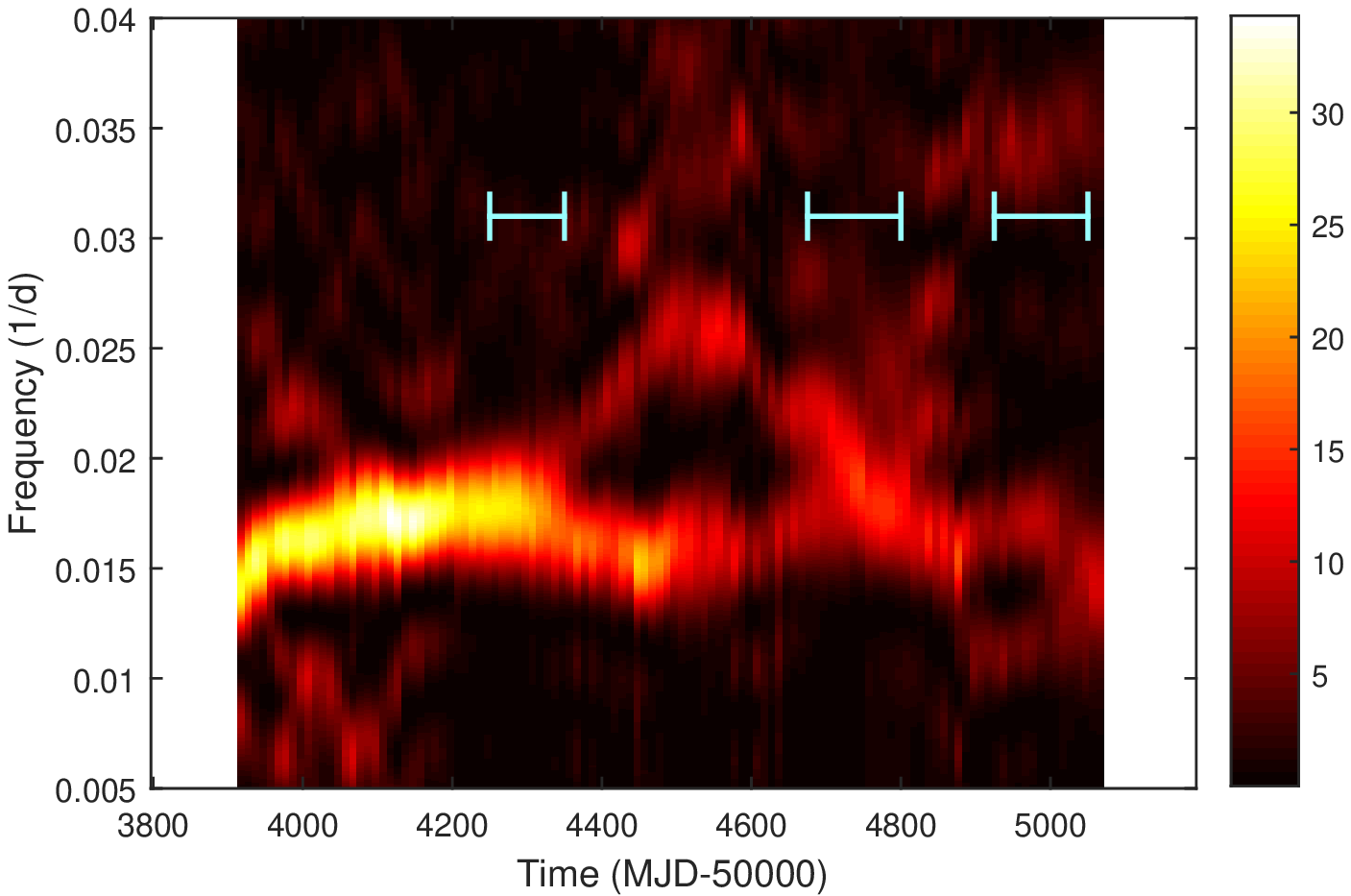,width=3.5in}
\epsfig{file=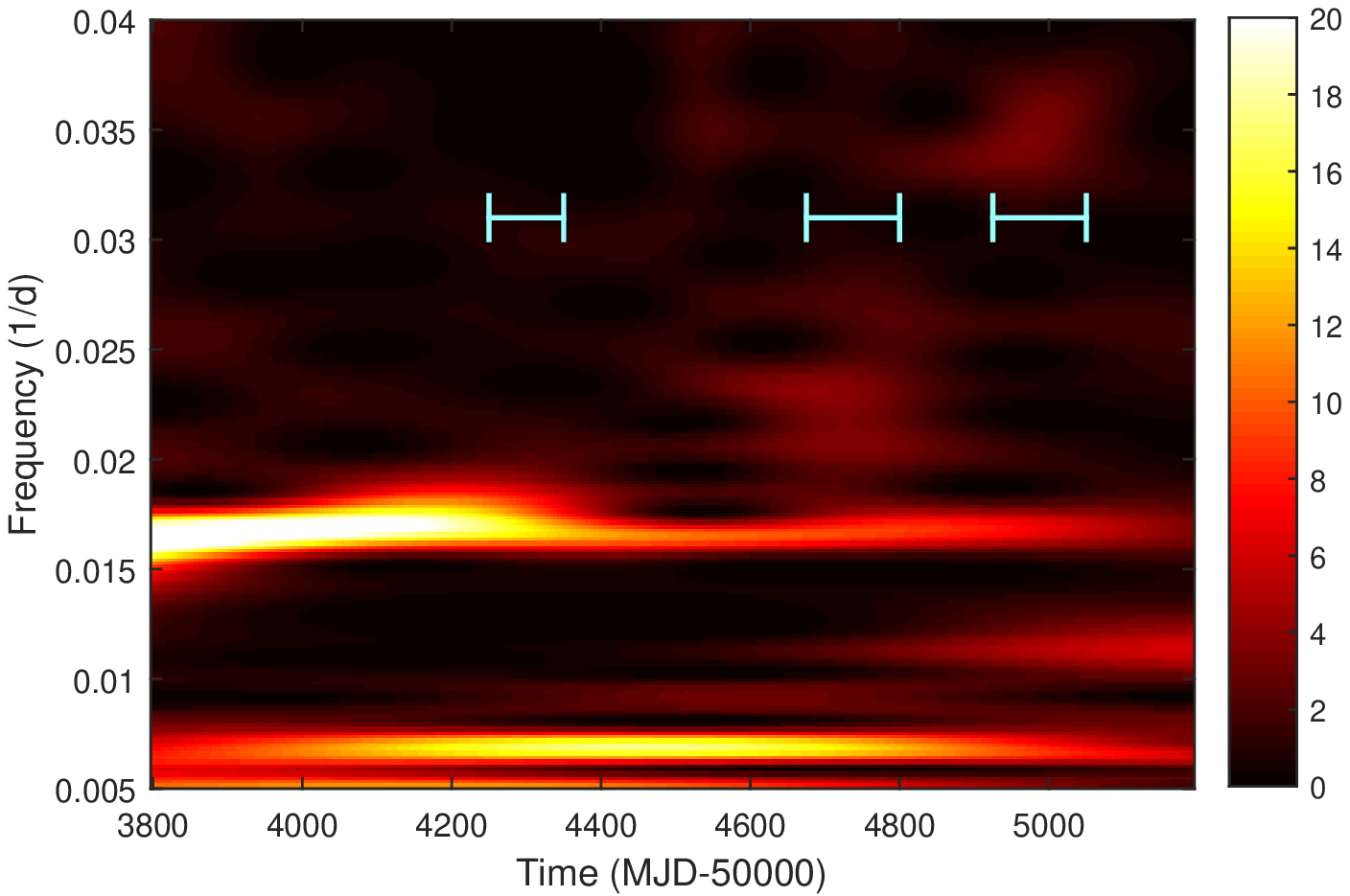,width=3.5in}
\epsfig{file=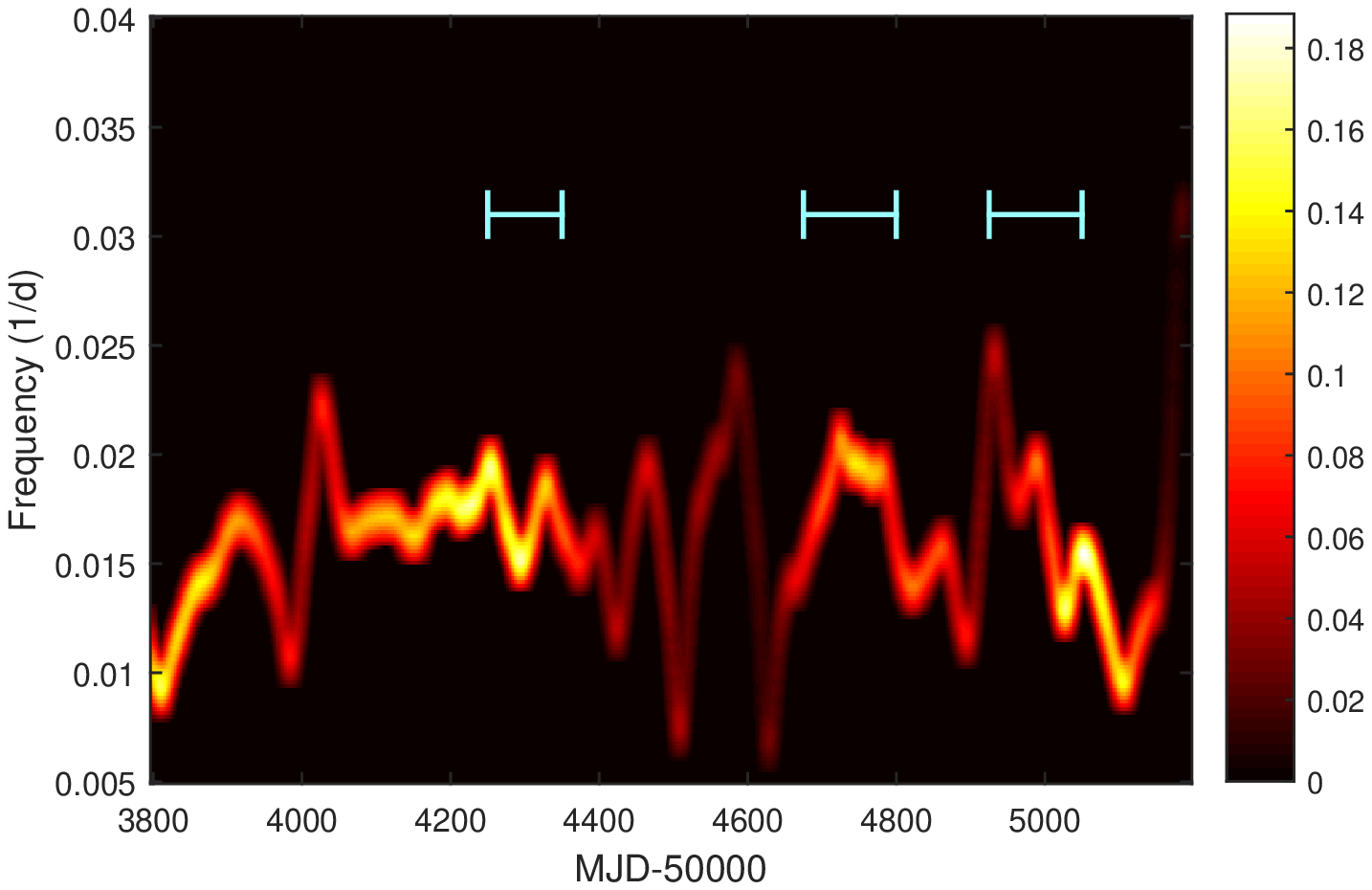,width=3.5in}
\caption{DPS (top), WWZ (middle), and HHT (bottom) spectra of \rxte/PCA observations of M82 X--1. The three flaring periods are marked with lines in the figures for reference. The signal at the frequency of 0.007 1/day seen in the WWZ map is caused by the similar duration of the three flares, leading an enhancement of the power after MJD 54,200. The same signal is not shown in the DPS because we have already filtered out the flaring effect with EEMD as described in Section 3.
\label{DPS_M82}}
\end{figure}

There are three flares (occurring at ~MJD 54250, 54700 and 55000) with durations comparable to the window size and may dominate the variability in the power spectrum, we used the ensemble empirical mode decomposition (EEMD; Wu \& Huang 2004) to locally filter out the long-term variability. We then applied the DPS on the EEMD high-pass filtered light curve with a window size of 240 days, which is roughly 4 cycles of the modulation period, and a moving step of 10 days.  The top panel of Figure \ref{DPS_M82} shows the DPS of the \rxte/PCA data of \ulx, where the period as well as the modulation amplitude indeed change with time. The change in modulation and amplitude of the periodicity might be related to the flares as indicated by the duration of flaring events in Figure \ref{DPS_M82}. The period of the main signal seems to gradually increase from $\sim$70 days ($\sim 0.0143$ 1/day) at the beginning of the observation, and then reaches a steady value of $\sim$58 days ($\sim 0.01724$ 1/day) after $\sim$MJD 54,100. After the end of the first X-ray flare ($\sim$ MJD 54,400), the modulation period returned to $\sim$ 61 days ($\sim 0.0164$ 1/day) but the significance drops dramatically.   Moreover, another signal with $f\sim$0.025 1/day seems appears, but this feature cannot be confirmed with the LSP.  Before MJD  54,400, the major WWZ signal is relatively stable, but its frequency is relatively lower at MJD 53,800 and becomes a little higher after MJD 54,000 (see middle panel of Figure \ref{DPS_M82}).  This change can also be interpreted by an increase in frequency.  After MJD 54,400, the strength of the main signal decreases as in the DPS, and some minor features start to appear in other frequency regions although their significance is weak.  The HHT can yield the instantaneous frequency and describe the frequency change in great details.  Bottom panel of Figure \ref{DPS_M82} shows the Hilbert spectrum of the main modulation component.  The increase in frequency at the beginning and the relatively stable modulation before MJD 54,400 can be clearly seen. After that, the frequency jumps dramatically and the amplitudes decrease significantly, indicating that the modulation is insignificant and the main periodic signal is not stable.

\begin{figure}
\epsfig{file=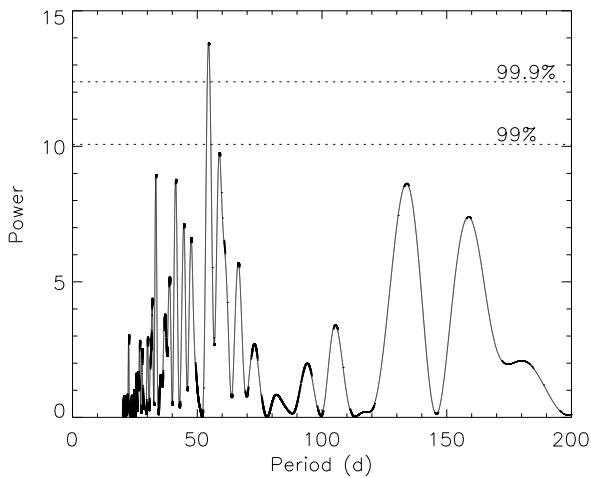,width=3.4in}
\caption{LSP of \swift\ XRT data with an extraction region of 18 arcsec in radius which includes all four ULXs.\label{lsp_swift}}
\end{figure}

\begin{figure*}
\epsfig{file=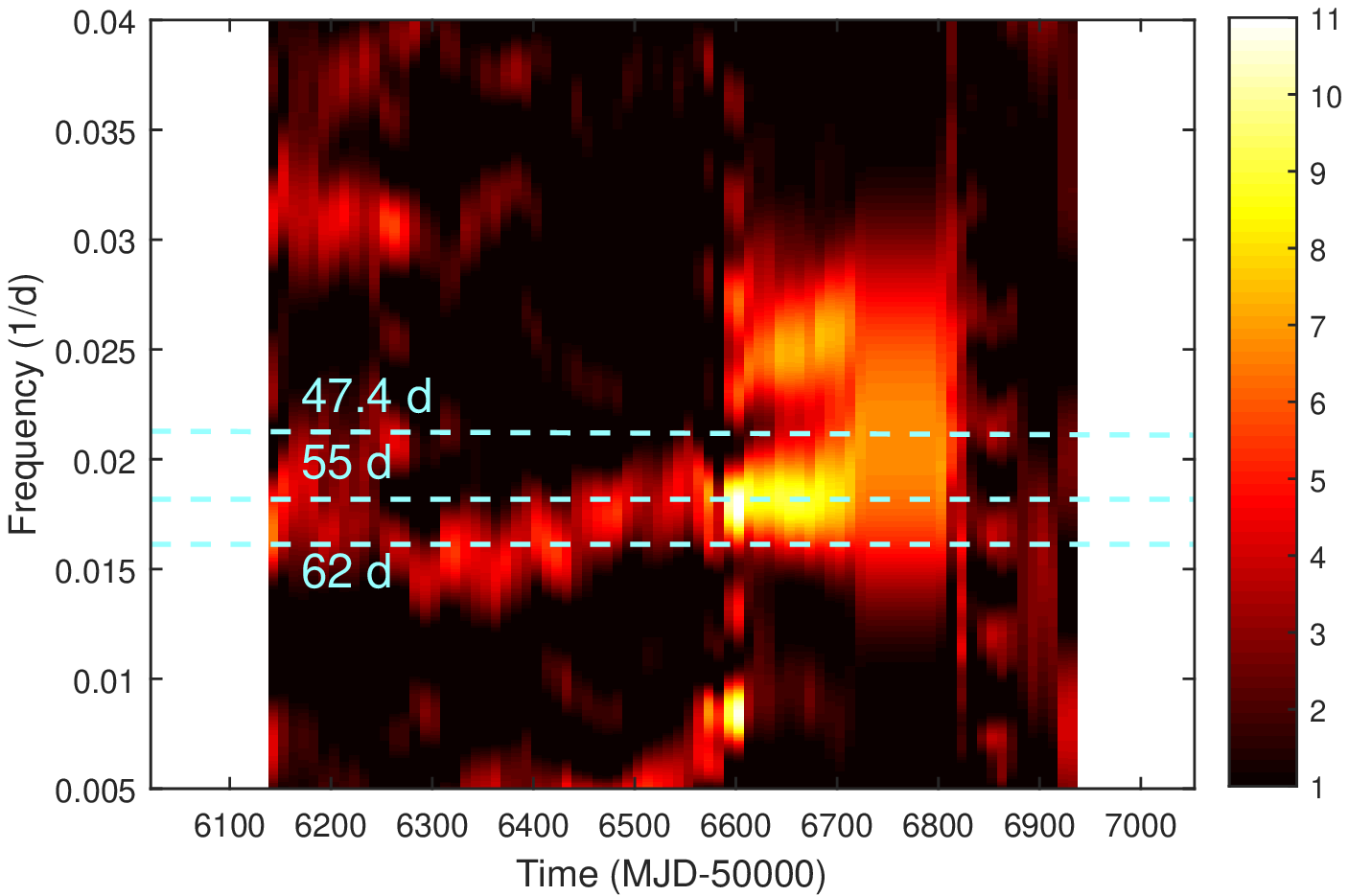,width=3.4in}
\epsfig{file=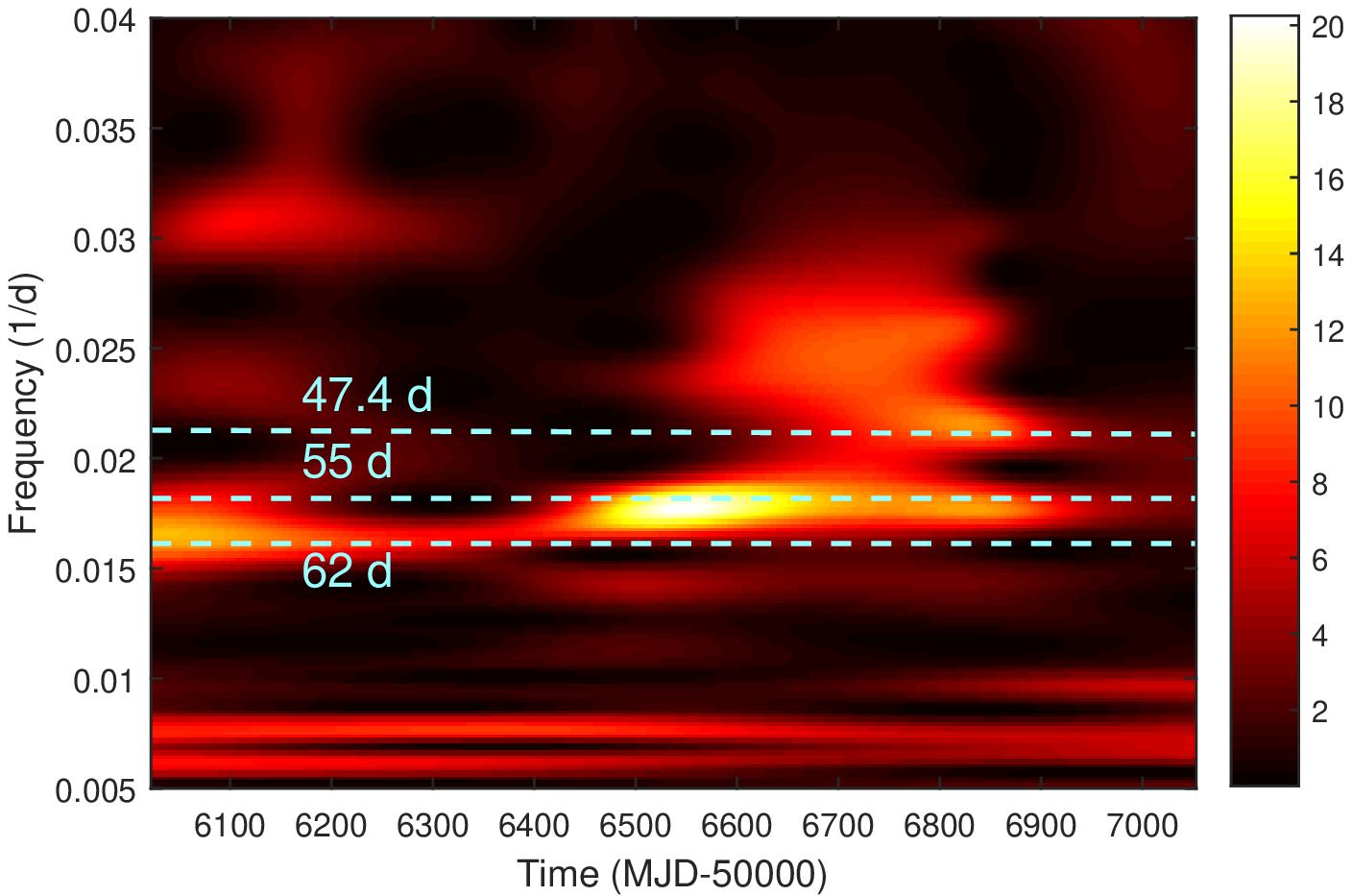,width=3.4in}
\caption{DPS (left) and WWZ (right) of \swift\ XRT data with an extraction region of 18 arcsec in radius. \label{DPS_swift}}
\end{figure*}

To investigate the modulations found in \rxte\ data, we repeated the LSP and the dynamic timing analysis with DPS and WWZ using \swift\ data.
Because there are two large data gaps at MJD 56,600 -- 56,670 and 56,770 -- 56,920 presented in the light curve, the HHT is no longer available in this examination.  
We first applied a relatively large extraction region consisting of \ulx\ as well as the three nearby ULXs to resemble the poor spatial resolution of \rxte. 
The LSP of the entire dataset is shown in Figure \ref{lsp_swift}. 
If we ignored the 70 observations made after the supernova SN 2014J, the result is consistent with that presented in Qiu et al. (2015).  
However, after adding those data points into our data set, the strongest signal is now at $P=54.5$ days ($\sim 0.01835$ 1/day) corresponding to a confidence level $> 99.9$\%.  
We further divided the data into two segments. 
For the data set obtained before MJD 56,500, the LSP shows a marginal detection at $P \sim 61.3$ days ($\sim 0.0163$ 1/day) with a confidence level of $\sim 99$\%.  
On the other hand, the power spectrum of data points obtained after MJD 56,650 shows two weak peaks ($\lesssim 99$\% significance level) located at $P\sim 47.4$ days ($\sim 0.021$ 1/day) and $P\sim 56.5$ days ($\sim 0.0177$ 1/day).
All the major peaks yielded before or after MJD 56,500 can be resolved in Figure \ref{lsp_swift}, but we cannot find their relations unless we perform the dynamical timing analysis.
According to the DPS shown in the left panel of Figure \ref{DPS_swift} with a windows size of 240 days and a moving step of 10 days, the major detected signal starts to appear with $P\sim61.3$ days (frequency $\sim0.0163$ 1/day) after MJD 56,100. In addition, the WWZ map (right panel of Figure 4) shows that the signal appears even earlier. This signal gradually shifts to $P\sim54-56$ days (frequency $\sim0.0178-0.0185$ 1/day), and after MJD 56,600, another sub-signal with $P\sim47.4$ days ($\sim0.021$ 1/day) starts to enhance and it seems to be split from the original signal. The major signal has the strongest power $P\sim55$ days ($\sim0.0182$ 1/day) from MJD 56,450--56,700 and its significance is strong enough to be resolved with the entire data set (as shown in Figure 3) or just the data collected after MJD 56,650. The peak at $P\sim61.3$ days and $\sim47.4$ days can also roughly be resolved in Figure 3 (static LSP) at a lower significance using the whole data set. We therefore do not favour to explain the major detected signal as a superposition of multiple signals. 
A totally consistent picture can also be obtained from the WWZ map as we demonstrate in the right panel of Figure \ref{DPS_swift}.  All these indicate that the modulations are not stable.

As suggested by Qiu et al. (2015), the modulations seen in \rxte\ are a combination of different signals from the four ULXs and therefore they employed a much smaller extraction region for \swift\ data. 
Following Qiu et al. (2015), we used a 4 arcsec radius extraction region centred at the \chandra\ positions of the ULXs (Chiang \& Kong 2011) and extracted the light curve for each of the four ULXs. 
For each source, we analysed with the LSP and DPS, and no statistically significant signal was detected among all of them.

\section{Discussion}

Part of the motivations of this study is to investigate the 62-day X-ray period claimed in previous \rxte\ and \swift\ observations (Pasham \& Strohmayer 2013; Qiu et al. 2015) with more sophisticated timing analysis and better datasets.
By using all the available \rxte\ and \swift\ data and employing a systematic dynamic analysis, we have confirmed the previous finding that the 62-day period in the nuclear region of M82 is not stable although it is statistically significant in the case of \rxte\ data (Pasham \& Strohmayer 2013; Qiu et al. 2015). The lack of stability can rule out the previous suggestion that the 62-day period is associated with the orbital period of \ulx\ (Kaaret \& Feng 2007). Furthermore, this modulation is contaminated by signals generated from the two nearby variable ULXs (X--2 and X--3) because of the poor spatial resolution of \rxte. For instance, M82 X--2 and X--3 can be as luminous as $10^{40}$ erg s$^{-1}$ (Kong et al. 2007), which are comparable to \ulx\ (Chiang \& Kong 2011). Following the method used in Qiu et al. (2015), we extracted the light curves of the four ULXs using \swift/XRT data with small extraction regions centred at each ULX. We did not find any significant signals in any of the four ULXs with the LSP. This contradicts the results in Qiu et al. (2015) where they found marginally significant signals at 55 days ($\sim 0.018$ 1/day) and 62 days ($\sim 0.016$ 1/day) for M82 X--2, X--3, and X--4. Nonetheless, they omitted the \swift\ data taken for the supernova SN 2014J in M82 by claiming that the data were contaminated by the supernova. However, SN 2014J has no known X-ray emission even with a 47 ks \chandra\ observation (Margutti et al. 2014). We visually inspected the \swift\ data and found no hint on SN 2014J as in Margutti et al. (2014). We therefore also included the 70 \swift\ observations taken between 2014 late-January and early-April. If the signals found in Qiu et al. (2015) are real, including this more frequent sampling dataset should increase the strength of the signals. However, we found negative results and this put the suggested modulations for any of the ULXs as seen in \swift\ datasets in question. We caution that this region is not well resolved with \swift\ and it is entirely possible that any real signals will be hidden by noise.

The main result of this paper is the use of the \chandra\ data to investigate all the proposed modulations near the centre of M82. The \chandra\ observatory provided an unprecedented spatial resolution to resolve the four ULXs near the nucleus of M82. 
By folding the data, we found evidence larger than $3\sigma$ significance level for a periodicity of $\sim 55$ days ($\sim 0.018$ 1/day) for M82 X--2. However, the null hypothesis probability to yield this detection on the LSP is only $\sim 0.06$ (i.e., less than $2\sigma$).
Because very few data points (only 12 useful observations) distributed in a very long time span and the signal may not be stable (as indicated in the \rxte\ and \swift\ data), the inconsistency between a sinusoidal fitting and the LSP is not unexpected. 
To further investigate the possible 55-day modulation, we also applied a bootstraping algorithm (e.g., Efron \& Tibshirani, 1993; Shao \& Tu, 1995) to examine the confidence range of the obtained detection for M82 X--2. Bootstraping is a statistical method specifically used for the cases of few samples, and it provides a systematic way to reject faked signals generated by random fluctuations of the data distribution and source flux. 
We assumed that the probability distribution of each simulation is uniformly distributed with a 90\% confidence interval of the observed data.
Under this assumption, we set an arbitrary mean value and a deviation to simulate different sets of light curve.  
The value of the flux in each simulated light curve has no relationship with each others, and we can check whether all these experiments can generate a periodicity of ~55-day or not. 
In our $10^4$ tests, we can always obtain a detection and a period of 55.5141-day ($\sim 0.018$ 1/day) with a standard deviation of 0.0043-day can be determined if we provided sufficient resolution to resolve the signal.  
Further high-resolution monitoring observations with \chandra\ will be crucial to confirm the modulation. It is worth noting that M82 X--2 is an accreting pulsar in a 2.5-day orbital period (Bachetti et al. 2014). If the 55-day modulation is real and is associated with M82 X--2, it will be the superorbital period of the system.

Superorbital periods are believed to be caused by irradiation-driven warping of accretion discs (Ogilvie \& Dubus 2001). The disc thus precesses and blocks the central compact object with a period longer than the orbital period of the system. Alternatively, the disc can also precess from tidal interaction (Whitehurst \& King 1991), while long-term modulation in the mass accretion rate can result in different X-ray states and intensities. For M82 X--2 with $M_X=1.4 M_\odot$ and $M_c \geq 5.2M_\odot$ (Bachetti et al. 2014), we can rule out the tidal interaction-driven disc precession scenario that requires $q=M_c/M_X < 0.25-0.33$ (Whitehurst \& King 1991). Furthermore, based on Chiang \& Kong (2011), the spectra of M82 X--2 are not changing over time although it shows irregular transient behaviour in terms of luminosity. Therefore it is unlikely that the X-ray modulations are due to variations in the accretion rate or changes in X-ray states.
Here, we argue that the 55-day modulation of M82 X--2 is likely due to an irradiation-driven warping disc. In Ogilvie \& Dubus (2001), they performed a stability analysis of accretion disc against the effect of irradiation in X-ray binaries in the context of the mass ratio between the companion and the compact object as well as the separation between the two objects.  Based on the separation between the two stars ($66.6\times10^{10}$ cm) and the mass ratio (Bachetti et al. 2014), we can put M82 X--2 in the steadily precessing warped disc region in the Figure 7 of Ogilvie \& Dubus (2001). That makes M82 X--2 similar to Her X--1, SS 433, and LMC X--4 which have relatively stable long-term X-ray modulations. Interestingly, the spin period and orbital period of M82 X--2 ($P_{spin}=1.37$ s, $P_{orbital}=2.5$ days) resemble to that of Her X--1 ($P_{spin}=1.24$ s, $P_{orbital}=1.7$ days) for which a 35-day superorbital period is found to be associated with a precessing warped disc. In addition, based on three-dimensional smoothed particle hydrodynamics simulations for the irradiation effect on an accretion disc, the mass ratio of the system will affect to what extent the disc will warp, tilt, and precess (Foulkes et al. 2006).
The mass ratio of M82 X--2 suggests that the entire accretion disc is tilted out of the orbital plane due to a strong twist developed in the disc (Foulkes et al. 2006,2010). In order to confirm the 55-day X-ray modulation for M82 X--2, a high resolution monitoring observation with \chandra\ is required in the future.

\section*{Acknowledgements} 
We thank the anonymous referee for the careful reading of the manuscript and for the useful comments, that helped improving the paper.
This work made use of data supplied by the UK Swift Science Data Centre at the University of Leicester. 
This project is supported by the Ministery of Science and Technology of the Republic of China (Taiwan) through grant 103-2628-M-007-003-MY3.

\label{lastpage}

\end{document}